\documentclass[prb,twocolumn,showpacs,amsmath,floatfix]{revtex4}
\usepackage{graphicx}
\usepackage{dcolumn}
\begin{document}
\newcommand{\be}{\begin{equation}}
\newcommand{\ee}{\end{equation}}
\newcommand{\ba}{\begin{eqnarray}}
\newcommand{\ea}{\end{eqnarray}}
\newcommand{\vk}{{\bf k}}
\newcommand{\vq}{{\bf q}}
\newcommand{\vp}{{\bf p}}
\newcommand{\vx}{{\bf x}}
\newcommand{\cA}{{\cal A}}

\title{ Effect of Hund coupling in the  one-dimensional SU(4) Hubbard  model}

\author{Hyun C. Lee,$^{1,2}$ Patrick Azaria,$^{3,4}$ and  Edouard Boulat$^{3,5}$}
\affiliation{$^1$ Department of Physics and Basic Science Research Institute, 
Sogang University, Seoul, 121-742, Korea}
\affiliation{$^2$ BK21 Physics Research Division and Institute of Basic Science,
 Department of Physics,Sung Kyun Kwan University, Suwon, 440-746, Korea}
\affiliation{$^3$ Laboratoire de Physique Th\'eorique des Liquides,
Universit\'e Pierre et Marie Curie, 4 Place Jussieu, 75252 Paris, France}
\affiliation{$^4$ CNRS-Centre National de le Recherche Scientifique, 
3, rue Michael-Ange, 75016 Paris, France}
\affiliation{$^5$ Department of Physics and Astronomy, Rutgers University, Piscataway, NJ 08855, USA}
\date{\today}
\begin{abstract}
The one-dimensional SU(4) Hubbard  model perturbed by Hund coupling is studied, 
away from half-filling, by means of  renormalization group and  bosonization methods. 
A spectral gap is always present in the spin-orbital sector irrespective
of the magnitude of the Coulomb repulsion. We further distinguish
between two qualitatively different regimes. At small Hund coupling,
we find that the  symmetry of  the system is dynamically enlarged  to SU(4)
at low energy with the result
of {\it coherent} spin-orbital excitations. When the charge sector is not gapped,
a superconducting instability is shown to exist. At large Hund coupling,
the  symmetry is no longer enlarged to SU(4) and the excitations in the spin sector
become  {\it incoherent}. Furthermore, the  superconductivity can be suppressed in favor
of the conventional charge density wave state.
\end{abstract}
\maketitle
\section{Introduction}
The interplay of  spin and  orbital degrees of freedom plays an important
role in diverse correlated electron systems. \cite{orbital}
Recently, the one-dimensional (1D) spin-orbital models have been studied intensively
motivated by the discovery of the quasi-1D spin-gapped materials,
$\mbox{Na}_2\mbox{Ti}_2\mbox{Sb}_2\mbox{O}$ and 
$\mbox{Na}_2\mbox{V}_2\mbox{O}_5$. \cite{experiment}
These  materials can be modeled by a quarter-filled {\it two}-band Hubbard model.
Even in this approximation, the situation is rather complex owing to the large
number of independent coupling constants in the problem.
 
A first attempt to understand
the effect of the high degeneracy induced by the orbital degrees of freedom was to consider
the situation  where both orbital and spin degrees of freedom play an identical role.
In this case, the {\it two}-band Hubbard model possesses a large symmetry and it 
becomes SU(4) symmetric in the spin-orbital sector and depends on only
one coupling: the Coulomb repulsion $U$. 
At {\it quarter}-filling it was found that spin-orbital and charge degrees
of freedom separate at low energy. 
The spin-orbital sector remains massless
for all values of  $U>0$ and displays quasi long-range antiferromagnetic order
with {\it three} critical modes whose dynamics is  described by a
SU(4) level 1 ($k$=1) Wess-Zumino-Witten (WZW) model. This result 
is in agreement with the exact result by Sutherland  for the SU(4) Heisenberg chain \cite{sutherland}.
What happens for the  U(1) charge excitations strongly
depends on $U$. It was found that a Mott-Hubbard transition from a massless
phase at small $U$ to an insulating phase at large $U$  takes place at a nonvanishing
critical value $U_c= 2.8t$.\cite{chargegap}
Of course, the SU(4) symmetry in the spin-orbital sector is not likely to be present in real materials,
so that a systematic study of the effects of  possible symmetry breaking
operators is necessary to  account for the experimental results eventually. 

One of the  simplest symmetry breaking is to break the SU(4) symmetry
down to ${\rm SU}(2)_{\rm spin}\times {\rm SU}(2)_{\rm orbital}$.  
A detailed renormalization group
study\cite{azaria, azaria2} revealed that the SU(4) symmetry is dynamically enlarged at low energy.
Furthermore, the  massless phase in the spin-orbital sector survives
in an extended region of coupling constant space. 
These  results were confirmed numerically.\cite{chigak}

In the present work we investigate a different symmetry breaking perturbation 
which is always present in real materials, the Hund coupling. 
We also consider the cases of general
fillings away from the  half-filled one.
In comparison with 
the spin-orbital model studied in Refs.[\onlinecite{azaria,azaria2}],
 the Hund coupling breaks the SU(4) symmetry further 
down to SU(2)$_{\rm spin} \times $U(1)$_{\rm orbital}$. As for our most important
result, we find that a spectral gap opens in the spin-orbital sector
for an {\it arbitrarily} small Hund perturbation. We further distinguish
between two qualitatively different regimes. At small Hund coupling, we find
that the SU(4) symmetry is dynamically enlarged at low energy like in the spin-orbital
model. We further  show that a superconducting (SC) instability is present in the charge sector. 
At large enough
Hund perturbation, the SU(4) symmetry is no longer fully enlarged. Instead, we find
a partially  SU(2) symmetry enlargement in the {\it orbital} sector. In this phase,
the SC instability may disappear in favor of the conventional charge density wave (CDW) instability.

This paper is organized as follows. In  Sec. II, we present our model and discuss
its symmetry properties. The renormalization group analysis at weak coupling is performed
in Sec. III,  where we also  discuss the physical properties in both spin-orbital and charge sector.
In Sec. IV,  we develop a complementary strong coupling approach for the {\it quarter}-filled 
case. We conclude this paper in Sec. V.

\section{Model Hamiltonian and its Symmetry Properties}
The Hamiltonian we consider  is the U(4) Hubbard model with a Hund coupling.\cite{ueda}
\be
H = H_0 + H_J,
\label{hamil}
\ee
with
\ba 
H_0 &=& \sum_{i \alpha \alpha^\prime \sigma} \Big(-
t^{\alpha \alpha^\prime}_{i,i+1}\,c^{\dag}_{i \alpha \sigma} c_{i+1 \alpha^\prime
\sigma}+{\rm H.c.}\Big) \nonumber \\
&+& \frac{U}{2}\,\sum_{i \alpha \alpha^\prime \sigma \sigma^\prime}\,
\Big[ n_{i \alpha \sigma} n_{i \alpha^\prime \sigma^\prime}(1-
\delta_{\alpha \alpha^\prime} \delta_{\sigma \sigma^\prime})\Big],
\label{hamilsu4}
\ea
 and
\be
H_J=-2 J \sum_i \, {\bf S}_{i1}\cdot{\bf S}_{i 2},
\label{hundcoupling} 
\ee
where the $c_{i \alpha \sigma}$ are  the electron operators at the  site $i$  
in  orbital $\alpha=(1,2)$ with  spin $\sigma$. In 
Eq. (\ref{hundcoupling}),
${\bf S}_{i \alpha}= \sum_{ \sigma \sigma'}\,
c^{\dag}_{i  \alpha\sigma}\, [\frac{\boldsymbol{\sigma}}{2}]_{\sigma
\sigma'}\,c_{i \alpha \sigma'}$,   denote  the
spin 1/2 operators   of electrons in both bands $\alpha=(1,2)$.
We further assume that the hopping is  diagonal in orbital space, i.e. 
$t^{\alpha \alpha^\prime}_{i,i+1}=t \delta_{\alpha \alpha^\prime}$, and that 
$U$ and $ J$ are positive. 

The symmetry properties of Eq. (\ref{hamil})  are most clearly seen by 
introducing the  U(1) charge $Q$ and the SU(4) spin-orbital generators 
${\cal T}^A$, $A=(1,\cdots,15)$, as follows:
\ba
Q &=& \sum_{i,\alpha \sigma}\,
c^{\dag}_{i \alpha \sigma} c_{i \alpha \sigma},\nonumber \\
{\cal T}^A &=& \sum_{i,\alpha^\prime \sigma^\prime \alpha \sigma}\,
c^{\dag}_{i,\alpha^\prime \sigma^\prime}\,
[M^A]^{\alpha^\prime \sigma^\prime}_{\alpha \sigma}\,
c_{i,\alpha \sigma},
\label{generator0}
\ea
where $M^A$ 
are the generators of SU(4) Lie algebra \cite{chigak,notations}.
A  convenient explicit realization of the $M^A$ is
\be
\label{generator}
\frac{1}{\sqrt{2}}\,(\frac{\sigma^a}{2})^{\sigma^\prime}_\sigma 
\delta^{\alpha^\prime}_\alpha,\,\,
\frac{1}{\sqrt{2}}\,\delta^{\sigma^\prime}_\sigma
(\frac{\tau^a}{2})^{\alpha^\prime}_\alpha, \,\,
\sqrt{2} (\frac{\sigma^a}{2})^{\sigma^\prime}_\sigma(\frac{\tau^b}{2})^{\alpha^\prime}_\alpha,
\ee  
where $\sigma^a$  and $ \tau^a$, $a=1,2,3$, 
are the Pauli matrices acting on spin and orbital
space, respectively. An appropriate labeling of the SU(4) generators in  Eq. (\ref{generator})
is as follows. To each SU(4) index $A=(1,\cdots, 15)$ we  associate a pair of indices, such that
$(a,b) \neq (0,0), (a,b=0,1,2,3) $ with the convention  that 
$\sigma^0=\tau^0={\rm Id}_2$. For an example, the first three generators of 
Eq. (\ref{generator}) can be alternatively expressed as $M^{(a,0)}$.

When $J=0$, using  Eqs. (\ref{generator0},\ref{generator}),  the Hamiltonian $H$
is clearly seen to commute with both $Q$ and all of ${\cal T}^A$'s,  thus it is 
U(4)=U(1)$_{{\rm charge}} \times $SU(4)$_{{\rm spin-orbital}}$ symmetric. 

The Hund coupling $H_J$ does not affect the charge sector but breaks 
the  SU(4)$_{{\rm spin-orbital}}$ symmetry. Indeed, when $J\neq 0$, in addition to
the obvious SU(2) invariance in spin space generated by $M^{(a,0)}$, 
$(a=1,2,3)$, $H$ is also invariant under
the  U(1)$_{\rm orbital}$ group  in orbital space
generated by $M^{(0,3)}$. Thus, the Hund coupling breaks the SU(4)$_{{\rm spin-orbital}}$ symmetry
down to SU(2)$_{\rm spin} \times $U(1)$_{\rm orbital}$. In comparison with the spin-orbital model 
studied in Refs.[\onlinecite{azaria,azaria2}] and Ref.[\onlinecite{chigak}], 
 the Hund term breaks the symmetry   
SU(2)$_{\rm orbital}$
further down to U(1)$_{\rm orbital}$.

\section{Renormalization Group Analysis  at weak coupling}
The effective low energy theory associated with the Hamiltonian (\ref{hamil}) is obtained 
 by taking the continuum limit in a standard way. 
At small enough $(U/t,J/t)$ and  at low energy, the  electron operators $c_{i \alpha \sigma}$
may be expanded near the two Fermi points $\pm k_F$:
\be
c_{i \alpha \sigma}= \sqrt{a_0}\; \Big[
e^{i k_F x} \psi_{R \alpha \sigma}(x)
+ e^{-i k_F x} \psi_{L \alpha \sigma}(x) \Big],
\label{contifermion}
\ee
where $a_0$ is the  lattice constant. In the continuum limit, the effective  Hamiltonian
is expressed naturally in terms of the chiral U(1) and SU(4) current densities: \cite{chargegap}
\ba
\label{current}
J_{L(R)}&=&\sum_{\alpha \sigma}\,
\psi^{\dag}_{L(R),\alpha \sigma} \psi_{L(R),\alpha \sigma},\nonumber \\
J_{L(R)}^\cA&=&\sum_{\alpha^\prime \sigma^\prime \alpha \sigma}\,
\psi^{\dag}_{L(R),\alpha^\prime \sigma^\prime}\,
[M^\cA]^{\alpha^\prime \sigma^\prime}_{\alpha \sigma}\,
\psi_{L(R),\alpha \sigma}.
\ea
After some algebra, discarding irrelevant operators,  we find that the charge
degrees of freedom decouples from the spin-orbital ones away from half-filling:
\be
 {\cal H} = {\cal H}_c + {\cal H}_{so},
\ee
where
\be
\label{hcharge}
{\cal H}_c = \int dx \left[ \, \frac{\pi v_{c}}{4}( J_{R}^2 + J_{L}^2) + g_c J_{R}J_{L}\right],
\ee
and
\ba
 \label{hspinorbit}
{\cal H}_{so} &=& \int dx \Bigg\{ \frac{2\pi v_{so}}{5}\sum_A(J^A_L J_L^A+J^A_R J_R^A)\nonumber \\
&-&\lambda_1   \sum_a \Big[ J_L^{(a0)} J_L^{(a0)} + J_R^{(a0)} J_R^{(a0)}  \Big]
\nonumber \\
&-& \lambda_2   \sum_a \Big[ J_L^{(a3)} J_L^{(a3)} + J_R^{(a3)} J_R^{(a3)}  \Big]
 \nonumber \\
&-& \tilde{g}_1  \sum_a  \Big[ J_L^{(a0)} J_R^{(a0)} \Big] \nonumber \\
&-&\tilde{g}_2   \sum_a \Big[ J_L^{(a,1)}J_R^{(a,1)} +
 J_L^{(a,2)}J_R^{(a,2)} \Big]\nonumber \\
 &-& \tilde{g}_3  \sum_a  \Big[ J_L^{(a3)} J_R^{(a3)} \Big]  \nonumber \\
 &-& \tilde{g}_4  
 \Big[ J_L^{(0,1)}J_R^{(0,1)}+ J_L^{(0,2)}J_R^{(0,2)} \Big] \nonumber \\
&-&\tilde{g}_5     J_L^{(0,3)}J_R^{(0,3)} \Bigg \}.
\ea
In Eqs. (\ref{hcharge},\ref{hspinorbit}),  
 $v_c= v_F(1  + 3Ua_0/2\pi v_F)$ and  $v_{so}= v_F(1 - Ua_0/2\pi v_F)$
are  the charge and spin-orbital velocities, where $v_F=2 ta_0 \sin k_F a_0$ is 
the Fermi velocity. We observe that all interactions of Eqs. (\ref{hcharge},\ref{hspinorbit})
are marginal and 
of the current-current type.\cite{note2} Therefore, the low energy physics will result
from a delicate balance among the different interaction terms in Eq. (\ref{hspinorbit}).
The  bare coupling constants in both charge and spin-orbital sectors  are expressed in term of $U$ and $J$
as follows: 
\be
g_c= \frac{3}{4} Ua_0, 
\ee
and
\ba
\label{initial}
 &\lambda_1=-\lambda_2=Ja_0, \;\; &\tilde{g}_1=2Ua_0+2Ja_0, \nonumber \\
 &  \tilde{g}_2=2Ua_0 +Ja_0, \quad &\tilde{g}_3=2Ua_0 - 2Ja_0,
 \nonumber \\
 &\tilde{g}_4=2Ua_0 -3Ja_0,  \;\;   &\tilde{g}_5=2Ua_0.
\ea

The effective  Hamiltonian in the charge sector [Eq. (\ref{hcharge})] is that of  Luttinger liquid:
\be
{\cal H}_c = \frac{v_c}{2} \int dx  \left[\frac{1}{K_c} (\partial_x \phi_c)^2 
+ K_c (\partial_x \theta_c)^2 \right],
\label{luttinger}
\ee
where $\phi_c = \phi_{cL} + \phi_{cR} $ and $\theta_c = \phi_{cL} - \phi_{cR}$
are  the charge boson field and its dual field, respectively. The chiral boson fields
$\phi_{c, L/R}$  are defined in terms of currents as follows:
\be
J_{L(R)}(x)=\sqrt{\frac{4}{\pi}}\;\partial_x \phi_{cL(R)}(x).
\ee
Therefore, the charge sector is  massless and the low energy properties
are determined by the nonuniversal  charge exponent $K_c$ which is given, at leading order  in $U$,
by:
\be
 K_c=\Big[1+\frac{3 U a_0}{\pi v_F}\Big]^{-1/2}< 1.
\label{kc}
\ee 
The charge velocity can be reexpressed as $v_c=v_F/K_c^2$.
The situation at hand is similar to what happens in the SU(4) Hubbard model at quarter filling.

The effective Hamiltonian in the spin-orbital sector is that of the 
SU(4)$_1$ Wess-Zumino-Novikov-Witten (WZWN) model with the central charge $c=3$, perturbed
by {\it marginal} interactions. This is similar to what happens in the spin-orbital model 
studied in  Refs.[\onlinecite{azaria,azaria2}] and Ref.[\onlinecite{chigak}].
 Due to the complexity
of the interaction pattern, namely the five coupling constants 
instead of three in the spin-orbital model,
the situation in the spin-orbital sector in the presence of a Hund term 
requires a  careful   analysis  of the renormalization group (RG) flow.  
Out of the seven coupling constants entering in Eq. (\ref{hspinorbit}), 
the $\lambda_1$ and $ \lambda_2$ terms  are purely   chiral and 
are not renormalized at   leading order. Furthermore, they   do not influence the scaling
of the  $\tilde{g}_i$'s. One-loop RG equations are easily found by current algebra 
method:\cite{affleck2,konik}
\ba
\label{rge}
\frac{d g_1}{d t}&=&-g_1^2-2 g_2^2-g_3^2, \nonumber \\
\frac{d g_2}{d t}&=&-2 g_1 g_2-g_2 g_5 - g_3 g_4, \nonumber \\
\frac{d g_3}{d t}&=&-2 g_1 g_3-2 g_2 g_4,  \nonumber \\
\frac{d g_4}{d t}&=&-3 g_2 g_3-g_4 g_5, \nonumber \\
\frac{d g_5}{d t}&=&-3 g_2^2-g_4^2,
\ea
where 
\be
g_i=\frac{\tilde{g}_i}{4\pi v_{so}},
\ee
and  $t=\ln L$ is  the RG time. 

We have performed a detailed numerical analysis of the RG flow associated with
Eqs. (\ref{rge}). In the following, we summarize  our results.

When $J=0$, the interaction is irrelevant for $U>0$ and
 all coupling constants  flow toward the SU(4)$_1$ fixed point at  $g_i^*=0$. There are no other
fixed points associated with Eq. (\ref{rge}). 
One of our most important results
is that   a nonzero value of the Hund coupling, $J\neq 0$,   destabilizes the SU(4)$_1$
fixed point and  drives the system toward  strong coupling. This indicates that a gap
opens in both spin and orbital sector with 
$M_{{\rm spin}} \sim M_{{\rm orbital}} \sim  \exp(-C/J)$, where $C$ is a positive constant of
order $t$.
The present situation
is completely different from the one encountered in  the spin-orbital model
 where the critical SU(4)$_1$ phase was {\it not} entirely
destabilized by the SU(2)$\times$SU(2) symmetry breaking perturbation.\cite{azaria}
Though a gap opens in the spectrum, the low energy effective theory
still depends on the relative magnitude, $\eta = J/2U$, between the Coulomb repulsion  and
the Hund  coupling. 
Indeed, one finds two qualitatively different behaviors  of the RG flow depending
on $\eta$. 

-\emph{The regime  A:  The \emph{SU(4)}  symmetric  regime.}  
The first is a  regime  with the enlarged SU(4) symmetry. This regime occurs for 
 $\eta \ll  1$ or $U  \gg  J$. 
This regime will be referred to as the regime A from now on.

With these initial conditions, though  all the coupling constants $g_i(t) \rightarrow \pm \infty$ 
when $t \rightarrow t^*$,  they asymptotically match the following particular RG-invariant ray:
\be
g_1=-g_2=-g_3=g_4=g_5  \rightarrow - \infty.
\label{restored}
\ee
On that ray, omitting chiral terms, one may write
Eq. (\ref{hspinorbit}) in an explicit SU(4) invariant form:
\be
{\cal H}_{so} = \sum_{A}\,\int dx \Big[ \frac{2\pi v_{s}^*}{5}(J^A_L J_L^A+J^A_R J_R^A) 
- \tilde{g}^* \, J^A_L J^A_R \Big ],
\label{su4}
\ee
where we have performed the  duality transformation for $a,b \neq 0$: 
\ba
J^{a,b}_R &\rightarrow&  - J^{a,b}_R, \nonumber \\
J^{a,b}_L &\rightarrow&   J^{a,b}_L,
\label{dual}
\ea
and $v_{s}^*$ is an effective spin-orbital velocity. In fact, there is a 
 velocity anisotropy in the model. We find,  however, that to leading order
in $J/t$ such an anisotropy of velocities scales to zero at low enough energy.\cite{notevelocity}
Thus, we find that the symmetry  is dynamically enlarged to SU(4)
 to the one-loop accuracy.
Of course the validity this result which relies on  the loop  expansion  may be  questioned, 
\cite{azaria,azaria3, balents} 
but it is reasonable to conjecture that the enlargement of the symmetry
 is likely to hold beyond the perturbation theory.
In any case  such an (enlarged) SU(4) symmetry is meant  to be  approximate 
in the  sense that  small corrections
to pure SU(4) behavior should be expected due to the neglected  irrelevant operators.

-\emph{The regime  B:  The \emph{SU(2)}$_{\rm orbital}$ enlarged regime}.
The second is a  regime B where SU(2)$_{\rm orbital}$ symmetry is {\it partially} 
enlarged [from U(1)].
This regime occurs for the   large Hund coupling $J \gg U$ or $\eta \gg 1$. 
For  $\eta  \gg  1$ we find that the SU(4) symmetry 
is no longer fully enlarged, and  instead we  observe
  a {\it partial}   SU(2)$_{\rm orbital}$ symmetry enlargement in the orbital sector.
With the initial conditions satisfying  $U \ll J$, 
the RG flows drive the coupling constants to   a regime where:
\be
- \infty \leftarrow g_4(t) <  g_5(t) \ll  g_2(t) <  g_3(t) \ll g_1(t) < 0.
\ee
In this regime the RG equations Eq. (\ref{rge}) can be approximately
decoupled. Indeed, at long RG time, the contributions of $g_2$ and $g_3$ to the RG equation for $g_4$ and $g_5$
can be neglected and one obtains:
\ba
\frac{d g_4}{d t}&\approx&-g_4 g_5,\nonumber \\
\frac{d g_5}{d t}&\approx&-g_4^2, 
\ea
which are nothing but the RG equations of the U(1) Thirring model in the orbital sector
with \emph{effective} initial conditions $|g_4(t)| >  |g_5(t)|$. 
In this regime, it is known
that the SU(2) symmetry is restored at larger RG time. Once the anisotropy between
$g_4$ and $g_5$ becomes small,  so does the anisotropy between $g_2$ and $g_3$ as can be seen
from the  equation 
\be
\frac{d (g_2-g_3)}{d t} \approx  g_4 ( g_2 -g_3), 
\ee
since  $(g_2-g_3) < 0$ and $g_4<0$. Therefore, in the strong coupling regime, the effective
 Hamiltonian approximately depends on {\it three} independent coupling constants:
\be
[ g_4  =  g_5 ] \ll [  g_2 = g_3 ]\ll g_1 <0.
\label{partialrestored}
\ee
With the above relation Eq. (\ref{partialrestored}), the interacting
part of Eq. (\ref{hspinorbit}) displays an SU(2)$_{\rm spin} \times$ SU(2)$_{\rm orbital}$ symmetry.
It is important to notice that no further symmetry restoration is expected 
since the coupling constant in the orbital sector $g_4$ is much larger than the one 
in the spin sector $g_1$. This behavior 
is in contrast with what happens in the A phase  for $\eta \sim 1$. 

The study of the physical properties near the boundary between the A and B regimes  
 is a nontrivial problem.  Whether they
are separated by a quantum phase transition point or they are smoothly connected by a crossover region 
can answered only by methods far beyond the perturbation theory.
This problem will be addressed elsewhere.
Within the one-loop accuracy, we find that the RG flow qualitatively changes from the A-type
to the B-type as  $\eta$ decreases below $\eta_0\
\sim 0.5$, which is a reasonable value,
but we were not able to conclude in favor of a quantum phase transition.
\subsection{Physical  Properties and Order Parameters}
\subsubsection{Spin-Orbital Sector}
In order to get some insights in the physical properties  of the low energy physics 
it is appropriate
to change the parametrizations of the fluctuating fields. 
We first notice that 
 the two sets of  SU(4)$_1$  currents,  $J^{(a,0)}_{\frac{R}{L}}$ and
$J^{(0,a)}_{R(L)}$, $a=(1,2,3)$, span two    spin and orbital SU(2) algebras. 
More precisely, they are  SU(2)$_{k=2}$ currents. This stems from the fact that the 
SU(4)$_{k=1}$ WZW model is equivalent to the sum of
two decoupled SU(2)$_{k=2}$ WZW model.\cite{affleck,zam}   As in Ref.[\onlinecite{azaria}]
 we shall take advantage
of the representation of the $SU(2)_{k=2}$ algebra in term of three (real) Majorana fermions
 \cite{zam,notation}:
\ba
\label{curr}
J^{(a,0)}_{R(L)}/\sqrt{2}&=&-\frac{i}{2}\,\epsilon^{abc}\,\xi^b_{s,R(L)}\,\xi^c_{s,R(L)},
\nonumber \\
J^{(0,a)}_{R(L)}/\sqrt{2}&=&-\frac{i}{2}\,\epsilon^{abc}\,\xi^b_{t,R(L)}\,\xi^c_{t,R(L)},
\nonumber \\
J^{(a,b)}_{R(L)} &=&- i\,  \xi^a_{s,R(L)}\xi^b_{t,R(L)},  \, \, (a,b) \neq 0,
\label{currmajo}
\ea
where $\xi_{s, R(L)}^a$ and  $ \xi_{t, R(L)}^a$, $a=(1,2,3)$, are  the 
Majorana fermions associated with 
the spin and  orbital degrees of freedom.
In term of these Majorana fermions the effective theories in both A and B regimes take a nice form.

In the regime A, the effective low energy Hamiltonian can be obtained from  Eq. (\ref{hspinorbit})
with the  condition $g_1=-g_2=-g_3=g_4=g_5=g < 0$ imposed:
\ba
\label{SO6}
{\cal H}&=&-i \frac{v^*_s}{2} \sum_a\,
\Big[ \xi^a_{s R} \partial_x \xi^a_{s R}-\xi^a_{s L} \partial_x \xi^a_{s L} \Big] \nonumber \\
&-&i \frac{v^*_s}{2} \sum_a\,
\Big[ \xi^a_{t R} \partial_x \xi^a_{t R}-\xi^a_{t L} \partial_x \xi^a_{t L} \Big] \nonumber \\ 
&-&g \, \Big[\sum_a (\kappa^a_s -  \kappa^a_t)\Big]^2,
\ea
where $\kappa^a_{s(t)} = \xi^a_{s(t) R}\xi^a_{s(t) L}$. The  Hamiltonian (\ref{SO6})
is easily seen to be SO(6) invariant upon a duality transformation in the orbital sector:
\ba 
\xi_{t, R}^a \rightarrow  -\xi_{t, R}^a,\quad
\xi_{t, L}^a \rightarrow  +\xi_{t, L}^a
\label{dualising} 
\ea
 which is the
equivalent of Eq. (\ref{dual}) when  Eq. (\ref{dual}) is expressed in terms of the Majorana fermions.
Under the  duality transformation Eq. (\ref{dualising}) , the  Hamiltonian (\ref{SO6}) becomes
that of the integrable SO(6) Gross-Neveu (GN) model \cite{gn1,gn2} which
has been first obtained in  Ref. [\onlinecite{azaria}] as the effective
low energy theory for the massive phase of the spin-orbital model. In this 
respect  we find that, though the Hund coupling Eq. (\ref{hundcoupling}) breaks the original  SU(4)
symmetry further than spin-orbital model, it is not  responsible 
for  the new low energy  physics  as far as $J$ is not too large in the spin-orbital sector. 
Therefore, many of the  conclusions drawn in Ref.[\onlinecite{azaria,azaria2,chigak}]
still hold for moderate values of $J$. In  particular, the excitation spectrum consists
of a  kink and an  anti-kink with mass $m$ along with
a  fundamental fermion of mass $\sqrt{2} m$. The existence 
of the fundamental fermion as a {\it stable} quasiparticle implies that the spin excitations
are {\it coherent}: a sharp resonance at $\omega=\sqrt{2}m$ is expected to appear 
in the dynamical structure factor, in particular, the $2k_F$ component of spin-spin 
correlation function. This can be checked by explicit calculation via order/disorder operator 
formalism of Ising model [Eqs. (\ref{ising1}, \ref{ising2})].

In the regime B, where $J \gg U$,  the effective  Hamiltonian is that of two {\it coupled} SO(3) GN
models, one in the spin sector and the other in orbital sector:
\ba
\label{largeJ}
{\cal H}_{so}&=&-i \frac{u_s}{2} \sum_a\,
\Big[ \xi^a_{s R} \partial_x \xi^a_{s R}-\xi^a_{s L} \partial_x \xi^a_{s L} \Big]
-g_1 \Big(\sum_a \kappa^a_s\Big)^2 \nonumber \\
&&-i \frac{u_t}{2} \sum_a\,
\Big[ \xi^a_{t R} \partial_x \xi^a_{t R}-\xi^a_{t L} \partial_x \xi^a_{t L} \Big]
-g_4 \Big(\sum_a \kappa^a_t\Big)^2 \nonumber \\
&&-g_2 \Big(\sum_a \kappa^a_s\Big) \Big(\sum_a \kappa^a_t\Big)
\label{SO3}
\ea
which is not integrable in general for arbitrary couplings $(g_1,g_2,g_4)$. However,
in the present case, where  $\eta\ll 1$ or $J \gg U$, the effective coupling constants
exhibit an interesting hierarchy:
\be
 |g_4|   \gg 
 |g_2|   \gg |g_1|.
\label{hiera}
\ee
As a consequence of the above hierarchy, one expects 
 the gap in the orbital sector ($ \sim  e^{-u_t /|g_4|}$) 
to be much larger than any other energy scale in the problem.
Therefore a mean-field decoupling of the interaction term in the Hamiltonian (\ref{SO3}) 
is sensible. We can carry out the  mean-field decoupling by introducing two Hubbard-Stranovich (HS) fields.
For that purpose we rewrite the interactions terms of Eq. (\ref{SO3}) as follows:
\ba
\label{hsdecouple}
& &g_1 \Big(\sum_a \kappa^a_s\Big)^2 +g_4 \Big(\sum_a \kappa^a_t\Big)^2 
+g_2 \Big(\sum_a \kappa^a_s\Big) \Big(\sum_a \kappa^a_t\Big)\nonumber \\
& &=\lambda_1 ( \Delta_t a_1 + \Delta_s b_1)^2 + \lambda_2 (\Delta_t a_2 + \Delta_s b_2)^2,
\ea
where the notations are 
\ba
\label{definitions}
\Delta_t &=& - i  \sum_a \xi^a_{t R} \xi^a_{t L},\;\; \Delta_s = - i  \sum_a \xi^a_{s R} \xi^a_{s L},\nonumber \\
\lambda_1& \sim & |g_4| + \frac{g_2^2}{4 |g_4|},\;\;
\lambda_2 \sim 2 |g_1| -\frac{g_2^2}{2 |g_4|}, \nonumber \\
(a_1,b_1)&=&( \frac{|g_4|}{\sqrt{(\frac{g_2}{2})^2+g_4^2}},\frac{|g_2|/2}{\sqrt{(\frac{g_2}{2})^2+g_4^2}}),\nonumber \\
(a_2,b_2)&=&(-\frac{|g_2|/2}{\sqrt{(\frac{g_2}{2})^2+g_4^2}},\frac{|g_4|}{\sqrt{(\frac{g_2}{2})^2+g_4^2}}).
\ea
In Eqs. (\ref{definitions}) the hierarchy Eq. (\ref{hiera}) was employed to simplify the expressions.
Cleary, $\lambda_1 \gg |\lambda_2|$.  Let us assume that $\lambda_2$ is positive. 
Next two HS fields, $\sigma, \zeta$,are introduced to decouple Eq. (\ref{hsdecouple}).
The resulting Hamiltonian in the action form can be written as
\ba
\label{actiondecouple}
S&=&\int dx d\tau \Big[ + \frac{1}{4\lambda_1}\,\sigma^2+ \frac{1}{4\lambda_2}\,\zeta^2 \\
&+&
\frac{1}{2}[\xi^a_{t R}\;\; \xi^a_{t L}]\,
\begin{bmatrix} 
\partial_\tau - i u_t \partial_x  &    i (\sigma a_1+\zeta a_2) \cr
- i (\sigma a_1+\zeta a_2) & \partial_\tau + i u_t \partial_x \cr
\end{bmatrix}
\begin{bmatrix}  \xi^a_{t R} \cr  \xi^a_{t L} \end{bmatrix} \nonumber \\
&+&
\frac{1}{2}[\xi^a_{s R}\;\; \xi^a_{s L}]\,
\begin{bmatrix} 
\partial_\tau - i u_s \partial_x  &    i (\sigma b_1+\zeta b_2) \cr
- i (\sigma b_1+\zeta b_2) & \partial_\tau + i u_s \partial_x \cr
\end{bmatrix}
\begin{bmatrix}  \xi^a_{s R} \cr  \xi^a_{s L} \end{bmatrix} \Big].\nonumber
\ea
Now the Majorana fermions can be integrated out exactly, and the effective action of $\sigma, \zeta$ is 
obtained. The saddle point approximation\cite{polyakov} of the effective action of $\sigma, \zeta$ gives
\ba
|\langle \sigma \rangle |& \sim & \Lambda_E e^{-u_t/|g_4|},\nonumber \\
\langle \zeta \rangle & \sim &\langle \sigma \rangle \frac{ |g_2|}{2 |g_4|}\,
\Big( \frac{\lambda_2}{\lambda_1} ( \frac{u_t}{u_s}-1)+
\frac{3}{\pi} \frac{\lambda_2}{u_s}\, \ln \frac{2 |g_4|}{|g_2|} \Big).
\ea
Note that $|\langle \sigma \rangle | \gg |\langle \zeta \rangle|$.
The saddle point values of $\Delta_t, \Delta_s$ are
\ba
\label{deltat}
\langle \Delta_t \rangle &\sim& \frac{\langle \sigma \rangle}{2 \lambda_1}, \\
\label{deltas}
\langle \Delta_s \rangle &\sim&  \frac{|g_2|}{2|g_4|}\,\frac{\langle \sigma \rangle}{2\lambda_1}\, 
\Big[\frac{u_t}{u_s} +  \frac{3}{\pi} \frac{\lambda_1}{u_s} \ln \frac{ |g_2|}{2 |g_4|} \Big ].
\ea
Note that  the factor of $\Big[\frac{u_t}{u_s} +  \frac{3}{\pi} \frac{\lambda_1}{u_s} \ln \frac{ |g_2|}{2 |g_4|} \Big ]$
in Eq. (\ref{deltas}) can be either positive or negative since $\ln \frac{ |g_2|}{2 |g_4|} < 0$.
In case the  factor is positive we have $\Delta_s \Delta_t >0$, while the opposite holds for  the negative factor.

Both the saddle point value and the fluctuation of $\zeta$ [note the factor $\zeta^2/\lambda_2 $] 
are very small compared to those of $\sigma$. Thus, due to the hierarchy Eq. (\ref{hiera}), the HS field
$\zeta$ can be neglected in the action Eq. (\ref{actiondecouple}).
The quantum fluctuations of $\sigma$ are large since $\lambda_1$ is large.
However, for the spin sector the fluctuations are suppressed by a factor of $b_1 \sim  |g_2|/2 |g_4|$ owing to 
the hierarchy [see the last line of Eq. (\ref{actiondecouple})]. Thus, we can take the saddle point value of 
$\sigma$ for the spin sector, while the full quantum fluctuations should be taken into account for the orbital sector.
Namely, a full integration over $\sigma$ is required for the orbital sector.
Then the  effective Hamiltonians in both spin and orbital sectors reduce to:
\be
{\cal H}_{\rm spin}=-i \frac{u_s}{2} \sum_a\,
\Big[ \xi^a_{s R} \partial_x \xi^a_{s R}-\xi^a_{s L} \partial_x \xi^a_{s L} \Big] 
- i m_s \Big(\sum_a \kappa^a_s \Big ),
\label{freemajo}
\ee
where $m_s = \sigma b_1  \sim |g_2| \Delta_t$, and
\be
{\cal H}_{\rm orbital}=-i \frac{u_t}{2} \sum_a\,
\Big[ \xi^a_{t R} \partial_x \xi^a_{t R}-\xi^a_{s L} \partial_x \xi^a_{s L} \Big] 
-  g_4 \Big(\sum_a \kappa^a_t \Big)^2.
\label{SO3gn}
\ee
In this limit, the spin excitations  consist of a  triplet
of  \emph{free} massive Majorana fermions with mass $m_s$  (or equivalently off-critical Ising models)
that span the spin one representation of SO(3)$_{\rm spin}$. This result can be understood 
qualitatively as follows. 
Since $J \gg U$,  the Hund coupling Eq. (\ref{hundcoupling}) dominate in Eq. (\ref{hamil})
and the two spin $1/2$ operators, ${\bf S}_{i1}$ and ${\bf S}_{i 2}$  are effectively
bound into a spin one state.
Thus, one  expects that in this limit the low energy sector of
Eq. (\ref{hamil}) may be identified with  that  of a doped  spin-one chain.\cite{frahm,dai}
This result will be recovered in the next section treating  the strong coupling limit. We also expect that in this limit
$J \gg U$, all single particle excitations will disappear of the
spectrum, since the strong Hund coupling tend to pair electrons.  
This is indeed the case : the kinks of the SO(6) GN model,
having the spin-orbital quantum numbers of the electron, vanish and
one is left in the lowest energy spectrum with massive Majorana
fermions, that is to say magnons.
The explicit calculation of (2 $k_F$ component of) spin-spin correlation functions 
using Ising model formalism [ Eqs. (\ref{ising1},\ref{ising2})]
shows that there is no sharp resonance. This implies that 
 the spin excitations are {\it incoherent}.

The orbital sector itself is described by an SO(3)$_{\rm orbital}$ GN model which
is integrable. Contrary to what happens in the spin sector, there are {\it no}
stable   (Majorana) fermions in the excitation spectrum\cite{gn2}.  
The kink and  anti-kink states with mass $m_t \sim e^{-u_t/|g_4|} \gg m_s$  exhaust the excitation
spectrum of SO(3) GN model.

We see that both A and B regimes differ in their spectral
properties, which is deeply related to the differences in the underlying symmetries.
Starting in the A phase and increasing the value of the Hund coupling $J$
we predict that, above a critical value $J_c$, the gap in the orbital sector $m_t$  becomes
much larger than the spin gap $m_s$:  the low energy excitations are 
exhausted  by  the spin excitations. 
Above $J_c$  fermionic excitations in the orbital sector  disappear and one is left
with  solely  kinks and  anti-kinks. This feature is reminiscent of a  decoherence phenomenon in 
orbital-like excitations. This prediction can be tested numerically.
\subsubsection{Charge Sector}
The fact that a gap opens  in the spin-orbital sector immediately suggests
the possibility of    charge density wave (CDW)  and 
superconducting (SC) 
instabilities. The corresponding   order parameters are given by:

\be
\hat{O}_{{\rm cdw}}= \sum_{a,\sigma} \psi^{\dagger}_{R a \sigma} \psi_{L a \sigma} + {\rm h.c}.
\label{cdw}
\ee
\be
\hat{O}_{{\rm sc,\pm}}=\psi_{R 1 \uparrow} \psi_{L 2 \downarrow} 
\pm \psi_{R 2 \uparrow} \psi_{L 1 \downarrow} + (R\leftrightarrow L),
\label{suprapm}
\ee
where $+/-$ stands for singulet/triplet SC.

For the discussion of the correlation functions of the above order parameters it is most 
convenient to use the Majorana fermion approach.\cite{azaria} 
To this end let us start with the  Abelian bosonization formulae:
\ba
\psi_{R(L), \alpha \sigma}=\frac{\kappa_{\alpha \sigma}}{\sqrt{2\pi a_0}}
\,e^{\pm i \sqrt{4\pi} \phi_{R(L),  \alpha \sigma}},
\ea
where  $\phi_{R(L) \alpha \sigma}$ are boson fields satisfying  
\be
[\phi_{R, \alpha \sigma},\phi_{L, \beta \sigma'}]
= i/4 \; \delta_{\alpha\beta}\delta_{ \sigma, \sigma'}.
\ee
$\kappa_{\alpha \sigma}$ are the Klein factors which enforce Fermi statistics: 
$$\{ \kappa_{\alpha \sigma} , \kappa_{\beta \sigma^\prime} \}=2 \delta_{\alpha \beta} 
\delta_{\sigma \sigma^\prime}.$$
A convenient choice of basis for the boson fields is  
\ba
\phi_{c,\frac{R}{L}}&=&\frac{1}{2} 
\Big( \phi_{\frac{R}{L},1 \uparrow}+ \phi_{\frac{R}{L},1 \downarrow}+ 
\phi_{\frac{R}{L},2 \uparrow}+ \phi_{\frac{R}{L},2 \downarrow} \Big)
\nonumber \\
\phi_{s,\frac{R}{L}}&=&\frac{1}{2} 
\Big( \phi_{\frac{R}{L},1 \uparrow}-\phi_{\frac{R}{L},1 \downarrow}+ \phi_{\frac{R}{L},2 \uparrow}
- \phi_{\frac{R}{L},2 \downarrow} \Big)
\nonumber \\
\phi_{f,\frac{R}{L}}&=&\frac{1}{2} 
\Big( \phi_{\frac{R}{L},1 \uparrow}+ \phi_{\frac{R}{L},1 \downarrow}
- \phi_{\frac{R}{L},2 \uparrow}- \phi_{\frac{R}{L},2 \downarrow} \Big)
\nonumber \\
\phi_{sf,\frac{R}{L}}&=&\frac{1}{2} 
\Big( \phi_{\frac{R}{L},1 \uparrow}- \phi_{\frac{R}{L},1 \downarrow}
- \phi_{\frac{R}{L},2 \uparrow}
+ \phi_{\frac{R}{L},2 \downarrow} \Big),
\ea
Refermionization formulae are given by:
\ba
\label{ising1}
\left(\frac{\xi^2 + i \xi^1}{\sqrt{2}}\right)_{R(L)} &=& 
\frac{\eta_{s}}{\sqrt{2\pi a_0}}\,e^{\pm i \sqrt{4\pi} \phi_{s,R(L)} }\nonumber \\
\left(\frac{\xi^5 + i \xi^4}{\sqrt{2}}\right)_{R(L)} &=& 
\frac{\eta_{f}}{\sqrt{2\pi a_0}}\,e^{\pm i \sqrt{4\pi} \phi_{f,R(L)}}\nonumber \\
\left(\frac{\xi^6 + i \xi^3}{\sqrt{2}}\right)_{R(L)} &=& 
\frac{\eta_{sf}}{\sqrt{2\pi a_0}}\,e^{\pm i \sqrt{4\pi} \phi_{sf,R(L)}},
\ea
where $\eta_{a}$ are new Klein factors. The spin and orbital Majorana  fermion triplets are
given by $\xi^a_s = (\xi^1,\xi^2,\xi^3)$ and 
$\xi^a_t = (\xi^4,\xi^5,\xi^6)$, respectively. 
In the Majorana fermion basis, both operators Eq. (\ref{cdw}) and Eq. (\ref{suprapm}) are nonlocal, 
while
they take local form  in terms of the order ($\sigma_a$) and the disorder ($\mu_a$) operators
of the six  underlying (off-critical) Ising models associated with the six Majorana fermions
 $(\xi^a_s, \xi^a_t) = (\xi^1, ...,\xi^6)$.
Using the correspondence: 
\ba
\label{ising2}
e^{ i \sqrt{\pi} \phi_{s} } \sim \mu_1 \mu_2 + i \sigma_1 \sigma_2,&    
e^{ i \sqrt{\pi} \theta_{s} } \sim \sigma_2 \mu_1 + i \mu_2 \sigma_1, \nonumber \\
e^{ i \sqrt{\pi} \phi_{f} } \sim \mu_4 \mu_5 + i \sigma_4 \sigma_5,& 
e^{ i \sqrt{\pi} \theta_{f} } \sim \sigma_5 \mu_4 + i \mu_5 \sigma_4, \nonumber \\
e^{ i \sqrt{\pi} \phi_{sf} } \sim \mu_3 \mu_6 + i \sigma_3 \sigma_6,&  
e^{ i \sqrt{\pi} \theta_{sf} } \sim \sigma_6 \mu_3 + i \mu_6 \sigma_3,
\ea
where $\phi=\phi_{L}+\phi_{R}$ and $\theta=\phi_{L}-\phi_{R}$ we find:
\ba
\hat{O}_{{\rm cdw}}&\sim &  \mu_1 \mu_2 \mu_3 \mu_4 \mu_5 \mu_6 \, \cos{\sqrt{\pi} \phi_{c}} 
\nonumber \\ 
&+& \sigma_1 \sigma_2\sigma_3\sigma_4 \sigma_5\sigma_6 \, \sin{\sqrt{\pi} \phi_{c}}.
\label{opsigmu1}
\ea 
\be
\hat{O}_{{\rm sc,+}}\sim e^{- i \sqrt{\pi} \theta_{c}} [\mu_1 \mu_2 + 
i \, \sigma_1 \sigma_2] [ \mu_3 \mu_4 \mu_5 \sigma_6 + \sigma_3 \sigma_4 \sigma_5 \mu_6].
\label{opsigmu2}
\ee
\be
\hat{O}_{{\rm sc,-}}\sim e^{- i \sqrt{\pi} \theta_{c}} [\mu_1 \mu_2 + 
i \, \sigma_1 \sigma_2] [ \sigma_3 \mu_4 \mu_5 \mu_6  -  \mu_3  \sigma_4 \sigma_5 \sigma_6].
\label{opsigmu}
\ee
We are now in a position to discuss the long distance properties of the correlation functions
 of the above order parameters. 

Consider first the regime A.
The spin-orbital dependent parts of Eq. (\ref{opsigmu1},\ref{opsigmu2},\ref{opsigmu}),
 which has scaling dimension $\frac{3}{4}$, 
 are expressed in terms  of products of six order and disorder operators $\sigma_a$ and $\mu_a$.
These $2^6$ operators constitute a basis for the  primary operators transforming in the spinor
representations of   SO(6)$_1$ WZW conformal
field theory. Among the spinorial primary operators, there are 
  two  SO(6) singlet primary operators : $\sigma_1 \sigma_2\sigma_3\sigma_4 \sigma_5\sigma_6$
and $\mu_1 \mu_2 \mu_3 \mu_4 \mu_5 \mu_6 $.
 At this point  it is worth stressing
that  the enlarged SO(6) symmetry present at low energy 
in the the regime A is  \emph{different} from the original SO(6) symmetry  of the noninteracting
theory. Two symmetries are related by the duality transformation Eq. (\ref{dualising})
in the orbital sector. Such a  transformation interchanges the order and the disorder
operators:
\ba
\sigma_a \leftrightarrow \mu_a, \; a=(4,5,6).
\ea
Consequently only  $\hat{O}_{{\rm sc,-}}$  contains  low energy SO(6) singlets
that can take a nonzero average value. Therefore, there exists   
quasi-long range triplet  superconducting  order.
\ba
\langle \hat{O}^{\dagger}_{{\rm sc,-}}(x,\tau) \hat{O}_{\rm sc,-}(0,0)\rangle  \sim
\frac{1}{\left(x^2+u_c^2\tau^2\right)^{1/4 K_c}}
\label{corrSC},
\ea
In contrast to the triplet superconducting  order, 
both CDW and singlet superconductivity  have short ranged correlations.
A similar analysis can be done when $J<0$ (antiferromagnetic). In this case the model  exhibits
 a singlet  superconducting instability rather than a triplet one.

In the  regime B  the situation is different. From
Eq. (\ref{deltat},\ref{deltas}) we find that $ \Delta_s   \Delta_t $ can be either positive or negative.
In case of $ \Delta_s   \Delta_t  > 0$,
  depending on the sign of $\Delta_t$ we have  at the mean field level
either:
\be 
\langle\mu_{a=1,2,3}\rangle =  \langle\mu_{a=4,5,6}\rangle \neq 0.
\ee
or
\be
 \langle\sigma_{a=1,2,3}\rangle= \langle\sigma_{a=4,5,6}\rangle\neq 0.
\ee
From Eqs. (\ref{opsigmu1},\ref{opsigmu2},\ref{opsigmu}), we conclude that 
 a CDW instability   is expected in this case.
In case of $ \Delta_s   \Delta_t  < 0$ we have either:
\be
\langle\mu_{a=1,2,3}\rangle =  \langle\sigma_{a=4,5,6}\rangle \neq 0.
\ee 
or 
\be
\langle\sigma_{a=1,2,3}\rangle =  \langle\mu_{a=4,5,6}\rangle \neq 0.
\ee 
Then a triplet superconductivity is expected in this case.

To summarize, we find that at weak coupling the Hund  perturbation  always opens
a gap in the spin-orbital sector. However, depending on the relative magnitude
of $U$ and $J$ we may distinguish between two qualitatively different regimes.
At small Hund coupling, $J < U$, the spin-orbital sector displays an effective low energy
with  enlarged SU(4) symmetry and there also exists  a triplet superconducting instability in the
charge sector. 
For  large Hund coupling, $J \gg U$, there is a partial SU(2)$_{\rm orbital}$
restoration, and depending on the relative magnitudes of parameters the superconducting instability may disappear in favor of
a CDW quasi-long range order. The way these two regimes are
connected is a nontrivial problem and requires a nonperturbative approach.

\section{Strong Coupling}
The situation  at large coupling depends on the filling as well as  the 
possible umklapp terms.\cite{chargegap}
For incommensurate  fillings all umklapp operators are   strongly oscillating 
and may be discarded at low energy. Consequently, the charge excitations are expected to remain
massless for all $U$ and $J$. 
For commensurate  filling, i.e. when  $n = p/q$, where $p$ and $q$ are co-prime numbers,
 the possible umklapp operators allowed by parity and translational invariance
are of the form:
\be
H_{{\rm umklapp}} =  \cos[4  q \sqrt{\pi} \phi_c].
\label{umklapp}
\ee
These operators represent the processes
that conserve total momentum up to an integer times the Fermi
momentum $k_F$ in the continuum limit, 
and they have the scaling dimensions:
\be
\label{scalingdimension}
\Delta_q= 4K_c  q^2.
\ee 
The equation (\ref{scalingdimension}) implies that the umklapp operators
are irrelevant as far as $K_c > 1/2  q^2$. 
From Eq. (\ref{kc}),
we find this is the case as far as  $U$ and $J$ are small enough.
However, as the coupling constants increase, we expect $K_c$
to decrease and possibly  to reach  the critical value $1/(2  q^2)$.
Below the critical value of $K_c$,  
the umklapp operator
Eq. (\ref{umklapp}) becomes relevant and a gap opens in the charge sector.
Thus, the mere existence of a Mott-Hubbard (MU) transition is related  to  the nonuniversal
dependence of $K_c$ on  the coupling constants $U$,$J$, and the filling $n$. The dependence is
not well-known in general.  At present, $K_c$ is only known at  {\it quarter}-filling and for 
 $J=0$.\cite{chargegap}
In  this case,  $K_c$   reaches its critical value
$K_c = 1/2$  at the value $U =2.8 t$, where  a MU transition
toward an insulating phase has been shown to occur.
What happens when $J\neq 0$ and for  other commensurate fillings
remains an open question. 

In this section we shall focus
on the  {\it quarter}-filled case and make a reasonable hypothesis that 
a  Mott transition still takes place in the presence of a Hund
term. Consequently we expect the model described by (\ref{hamil}) to
be an insulator  for  large $U$ and $J$ when  $n=1$. For this particular filling 
the  strong coupling regime  is best achieved by going to the Heisenberg limit.
As shown by Arovas and Auerbach \cite{arovas} all of the relevant low energy states
at strong coupling regimes are given by:
\ba
\bar{u}_0&:&\frac{1}{\sqrt{2}}\,(c^{\dag}_{1 \uparrow} c^{\dag}_{2 \downarrow}-c^{\dag}_{1 \downarrow}
 c^{\dag}_{2 \uparrow}) |0 \rangle,
\nonumber  \\
\bar{u}_1&:& c^{\dag}_{1 \uparrow} c^{\dag}_{2 \uparrow} |0 \rangle,
\,\frac{1}{\sqrt{2}}(c^{\dag}_{1 \uparrow} c^{\dag}_{2 \downarrow}+
c^{\dag}_{1 \downarrow} c^{\dag}_{2 \uparrow}) |0 \rangle,
 \,c^{\dag}_{1 \downarrow} c^{\dag}_{2 \downarrow}  |0 \rangle,\nonumber \\
\bar{u}_2&:& c^{\dag}_{1 \uparrow} c^{\dag}_{1 \downarrow} 
|0 \rangle,\;c^{\dag}_{2 \uparrow} c^{\dag}_{12\downarrow} |0 \rangle.
\ea
The above states represent inter-orbital spin-singlet, inter-orbital spin-triplet, 
and intra-orbital spin-singlet states. The energy of each state is given by:
\be
\bar{u}_0=U+\frac{3}{2} J,\;\;
\bar{u}_1=U-\frac{1}{2}J,\;\;
\bar{u}_2=U.
\label{energies} 
\ee
The effective strong coupling   Hamiltonian  depends  crucially  on the value of $\bar{u}_1=U-J/2$.

\subsection{ $U \gg  J/2 \gg t$}
In this case all the energies in Eq. (\ref{energies}) are positive and we can employ
the results derived by Arovas and Auerbach.\cite{arovas}
They found that  the effective  Hamiltonian
is given by a generalization of the spin-orbital model:
\ba
\label{AAH}
{\cal H}_{{\rm eff}}&=& \sum_i \Big[ A_1 {\bf S}_i \cdot  {\bf S}_{i+1} + 
A_2 {\bf T}_i \cdot {\bf T}_{i+1}+
A_3 T^z_i T^z_{i+1}  \nonumber \\
&+&A_4 {\bf S}_i \cdot  {\bf S}_{i+1} {\bf T}_i \cdot {\bf T}_{i+1}
+A_5 T^z_i T^z_{i+1}  {\bf S}_i \cdot  {\bf S}_{i+1} \Big],
\ea
where
\ba
{\bf S}_{i}&=& \sum_{ \alpha \sigma \sigma'}\,
c^{\dag}_{i  \alpha\sigma}\, \left[\frac{\boldsymbol \sigma}{2}\right]_{\sigma
\sigma'}\,c_{i \alpha \sigma'}, \nonumber \\
{\bf T}_{i}&=& \sum_{ \sigma \alpha \alpha'}\,
c^{\dag}_{i \alpha \sigma}\, \left[\frac{\boldsymbol \sigma}{2}\right]_{\alpha
\alpha'}\,c_{i   \alpha'\sigma},
\ea
and
\ba
A_1&=& \frac{ 2 t^2}{U}\,\frac{(1-3 \eta^2)}{(1-\eta)(1+ 3 \eta)},\nonumber \\
A_2&=&  \frac{ 2 t^2}{U}\,\frac{(1+5 \eta)}{(1-\eta)(1+ 3 \eta)},\nonumber \\
A_3&=&- \frac{ 2 t^2}{U}\,\frac{ 3 \eta}{(1+3 \eta)},\nonumber \\
A_4&=& \frac{ 2 t^2}{U}\,\frac{4 (1+\eta)}{(1-\eta)(1+ 3 \eta)},\nonumber \\
A_5&=&-4 A_3,
\ea
 with $\eta \equiv \frac{J}{2 U} \ll 1$.
When $\eta=0$, the  Hamiltonian Eq. (\ref{AAH}) is the SU(4) invariant 
antiferromagnetic Heisenberg model studied by
Sutherland \cite{sutherland}. 
Expanding the Hamiltonian Eq. (\ref{AAH})  with respect to  the SU(4) symmetric point we obtain
\ba
{\cal H}_{{\rm eff}}&\sim & {\cal  H}_{SU(4)}  \nonumber \\
&+& \frac{ 2 t^2 \eta}{U}\, \sum_i \Big[ -2 {\bf S}_i \cdot {\bf S}_{i+1} 
+3 (T^x_i T^x_{i+1}
+T^y_i T^y_{i+1}) \nonumber \\
&-&4  {\bf S}_i \cdot {\bf S}_{i+1}  {\bf 
T}_i \cdot {\bf T}_{i+1}+12 T^z_i T^z_{i+1} {\bf S}_i \cdot {\bf S}_{i+1} \Big] .
\label{heffstrong}
\ea
Following Ref. [\onlinecite{azaria}]  we find that  in  the continuum limit the effective  Hamiltonian
Eq. (\ref{heffstrong}) is identified with that of Eq. (\ref{hspinorbit}).
When   $\eta <1$, a RG analysis 
reveals the same enlargement of SO(6) symmetry as in the A regime  of the weak coupling case.
Thus,  one may conclude that the SO(6) symmetric A regime  extends
from weak to strong couplings for small  enough  Hund interaction.

\subsection{$t \ll U \ll  J/2$}
In this case $\bar{u}_1$ becomes negative and the strong coupling approach
 developed by Arovas and Auerbach does not apply. However, another strong coupling
 expansion is sensible when  $U \ll  J/2$. Indeed, in this limit
the ground state consists of the local spin triplets which  contain
{\it two} electrons per site and spontaneously break translational
invariance. Assuming  that the spin triplets are located on  even sites (such that  odd sites are empty)
we  can find an effective interaction  between  the  local spin triplets.
The effective interaction can be determined from the strong coupling expansion to the order of $t^4$
in a straightforward way.
Let us denote  the spin triplet at the $2j$-th site by
${\bf I}_{2j}$,  which is  a $S=1$ spin operator.
By solving the associated two-site problem up to the order  of $t^4$, 
the following effective  Hamiltonian can be obtained: 
\be
\label{afhamil}
{\cal H}_{{\rm eff}}^{(4)}=\sum_j \Big[  K_2 ({\bf I}_{2 j} \cdot {\bf I}_{2j+2})^2+
K_1 {\bf I}_{2 j} \cdot {\bf I}_{2j+2} \Big ],
\ee
with
\ba
K_1&=& \frac{8 t^4}{5 J^3} \frac{ 7-31 \epsilon/2+2 
\epsilon^2+\epsilon^3/2}{(1-\epsilon)^3(1-\epsilon/5)(1-\epsilon^2/4)},
\nonumber \\
K_2&=&-\frac{24 t^4}{5 J^3} \frac{1}{(1-\epsilon/5)(1-\epsilon)^2},
\ea
where $\epsilon =  \frac{2 U}{J} \ll 1$.
The  Hamiltonian  Eq. (\ref{afhamil}) is that of the  {\it antiferromagnetic}  Heisenberg
spin 1 model with exchange   $K_1$ and a biquadratic coupling $K_2 < 0$. In this regime
of coupling constants, the excitations are massive and consist of a triplet of Majorana
fermion. \cite{nersesyan}
 It is however hazardous to conclude that the
Haldane magnons constitute the low energy excitations in this strong
Hund coupling regime. Indeed, the groundstate for the $t=0, U\ll J$
limit of the model is doubly degenerate for the local spin triplets
can be located on even or odd sites. In addition to the Haldane
magnons of (\ref{afhamil}) we thus have to take into account the kinks
that connect these two groundstates.  Note that these kinks carry
integer spin, since excitations built out of single electrons have
very high energy in this $U\ll J$ regime.

A rough estimate of the energy of kink excitations can be obtained in
a static approximation.  Consider a pined kink-antikink pair on top of
the groundstate of (\ref{afhamil}), located at sites $p$ and $q$ ($p,q$
beeing odd integers), that is to say that the local spin triplets are
located on even sites $2i$ for $2i<p$ and $2i>q$ and on odd sites
$2i+1$ for $p \leq 2i+1<q$. When the $t\neq 0$ perturbation is
included, this will result in the effective hamiltonian
(\ref{afhamil}) with modified exchange $K_{1,2}$ on the bounds $p$ and
$q$. The static approximation is thus equivalent to two bound defects,
one weak and one strong (with transparent notations, $K_{1,2}(p)\gg
K_{1,2}\gg K_{1,2}(q)$ for the particular configuration we have
chosen). The effect of bound defects has been studied in the spin 1
pure Heisenberg chain (without biquadratic exchange), which has the
same low energy physics as (\ref{afhamil}). The defects there lead to
the apparition of a triplet excitation \emph{inside} the Haldane
gap.\cite{sorensen} It is more than likely that beyond this static
approximation, the kinks will form a band of spin 1 excitations once
kinetic energy is included, as occurs for holes in the AKLT
model.\cite{zhang}

While this problem would clearly require a more careful analysis,
which goes well beyond the scope of this paper, this simple static
picture indicates that the lowest energy excitations will be massive
spin 1 kinks. In any case, the low energy spectrum is exhausted by
massive magnons.  The effective continuum theory in regime B at weak
coupling $t \gg J \gg U$ describes also massive spin 1 particles ; it
is thus tempting to postulate a continuity from weak coupling to
strong coupling for the low energy excitations.

\subsection{$U \sim J/2 \gg t$}
In this case the local spin triplet {\it two particle} states have very low energy 
and they will mix with other  states with {\it one particle per site}
in the first order of hopping $t$.  These states should be diagonalized
 first before taking into account the higher order perturbation in $t$.
The detailed study of this regime is beyond the scope of this paper.

\section{Summary}
We have studied the one-dimensional  SU(4) Hubbard  model away from 
half-filling perturbed by a Hund coupling $J$ in both the weak and strong 
coupling limits.
The Hund coupling turns out to be relevant irrespective of the 
short ranged  Coulomb repulsion $U$ and is responsible for the opening of a 
spectral gap in the spin-orbital sector. We found two qualitatively
different regimes depending on the relative strength between $J$ and $U$.

\begin{figure}[ht]
\begin{center}
\scalebox{0.4}{\includegraphics{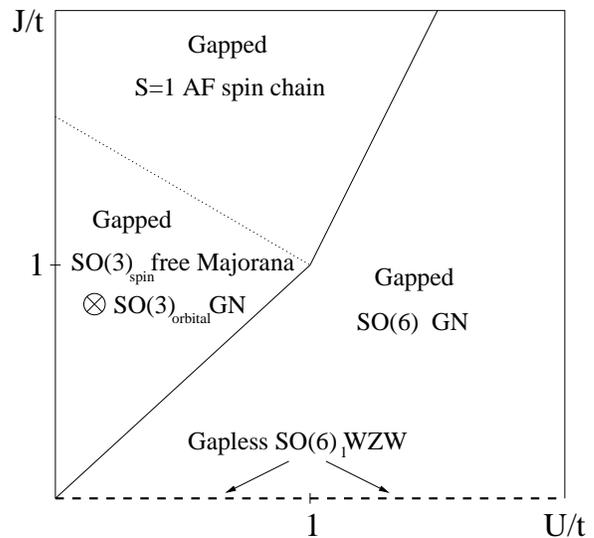}}
\caption{The phase diagram for the 
 spin-orbital degrees of freedom.
The symmetry as well as the effective model at the strong coupling (low energy) regime are indicated.  
Boundaries represent  smooth crossover
rather than critical quantum phase transitions.}
\label{spectrereseau}
\end{center}
\end{figure}

For small enough Hund coupling $J$, we found that the  symmetry
is dynamically enlarged to SU(4) at low energy. 
In this limit, the the spin-orbital degrees of freedom are 
described  the integrable  SO(6) Gross-Neveu  model 
and the excitations are found to be {\it coherent}. In this regime,
we found a superconducting instability when the charge  excitations are not gapped.
For large  Hund coupling the SU(4) symmetry is no longer enlarged.
Instead, we find that the orbital degrees of freedom decouples at high energy
and the low energy excitations lie in the spin sector and consist
of the  three  free massive Majorana fermions. As a result, the spin excitations
become {\it incoherent}. Our results  are summarized  in Fig. \ref{spectrereseau}.

\begin{acknowledgments}
P. Azaria and E. Boulat want to acknowledge P. Lecheminant
for valuable discussions and suggestions.
H.C. Lee is grateful to Prof. K. Ueda for the suggestion of this problem.
This work was partially supported by the Korea Science and Engineering
Foundation (KOSEF) through the grant No. 1999-2-11400-005-5, and by the 
Ministry of Education through Brain Korea 21 SNU-SKKU Program.
\end{acknowledgments}



\begin{thebibliography}{}
\bibitem{orbital} T. Tokura and N. Nagaosa, Science {\bf 288}, 462 (2000).
\bibitem{experiment}  E. Axtell {\it el al.}, J. Solid State Chem. {\bf 134},
423 (1997); M. Isobe and Y. Ueda, J. Phys. Soc. Jpn. {\bf 65}, 1178 (1996).
\bibitem{sutherland} B. Sutherland, Phys. Rev. B {\bf 12}, 3795 (1975).
\bibitem{chargegap} P. Assaraf, P. Azaria, M. Caffarel, and P. Lecheminant,
Phys. Rev. B {\bf 60}, 2299 (1999).
\bibitem{azaria} P. Azaria, A. O. Gogolin, P. Lecheminant, and A. A. Nersesyan,
Phys. Rev. Lett. {\bf 83}, 624 (1999).
\bibitem{azaria2}
P. Azaria, E. Boulat, and P. Lecheminant, Phys. Rev. B {\bf 61}, 12112 (2000).
\bibitem{chigak} C. Itoi, S. Qin, and I. Affleck, Phys. Rev. B {\bf 61}, 6747
(2000).
\bibitem{ueda} Y. Yamashita, N. Shibata, and K. Ueda, Phys. Rev. B {\bf 58}, 
9114 (1998).
\bibitem{notations} From now on the calligraphic letters denote the SU(4) 
generators.
\bibitem{note2} Note the differences in the sign conventions of coupling constants 
between ours and those of Konik {\it et al.}\cite{konik}.
\bibitem{affleck2} I. Affleck, in {\it Fields, Strings, and Critical Phenomena,
Les Houches 1988}, edited by E. Brezin and J. Zinn-Justin (North-Holland, 
Amsterdam, 1990).
\bibitem{konik} R. Konik, H. Saleur, A.W.W. Ludwig, cond-mat/0009166.
\bibitem{notevelocity} Velocity anisotropy does renormalize at one
loop order, provided one performs the perturbative expansion around the
\emph{non} Lorentz invariant fixed describing the free theory with
 velocity anisotropy. This
is best formulated in term of Majorana fields (see Eq. (\ref{currmajo})),
which acquire three different velocities : $v_s$ for $\xi_s^a$, 
$v_t$ for  $\xi_t^4,\xi_t^5$ and $v_6$ for  $\xi_t^6$. Our one loop
calculation shows that in the far infrared, these velocities 
converge to the one loop RG invariant 
$v_s^*=\left(v_s^3v_t^2v_6\right)^{1/6}$. A very similar situation
is discussed in detail in Section III A of Ref.[\onlinecite{azaria2}].
\bibitem{azaria3} P. Azaria, P. Lecheminant, A. Tsvelik, cond-mat/9806099.
\bibitem{balents} L. Balents and M. P. A. Fisher, Phys. Rev. B {\bf 53}, 12133.
\bibitem{affleck} I. Affleck, Nucl. Phys. B {\bf 265}, 409 (1985).
\bibitem{zam} A. B. Zamolodchikov and V. A. Fateev, Sov. Jour. Part. Nucl. {\bf 43}, 657
(1986).
\bibitem{notation} We follow the notations of P. Azaria {\it et al.}\cite{azaria,azaria2}.
\bibitem{gn1} D. Gross and A. Neveu, Phys. Rev. D {\bf 10}, 3235 (1974);
R. Dashen, B. Hasslacher, and A. Neveu, Phys. Rev. D {\bf 12}, 2443 (1975).
\bibitem{gn2} A. B.  Zamolodchikov and Al.  B. Zamolodchikov, Ann. Phys. (N. Y.) {\bf 120},
253 (1979); R. Shankar and E. Witten, Nucl. Phys. B {\bf 141}, 349 (1978);
M. Karowski and H. Thun, Nucl. Phys. B {\bf 190}, 61 (1981).
\bibitem{polyakov} A. M. Polyakov, {\it Gauge Fields and Strings} (Harwood, Chur, Switzerland, 1987).
\bibitem{frahm} H. Frahm, M. Pfannm\"{u}ller, and A. Tsvelik, Phys. Rev. Lett.  {\bf 81}, 2116 (1998).
\bibitem{dai} X. Dai, B. Chen, and Z. Su,  Phys. Rev. B, {\bf 59}, 11801 (1999).
\bibitem{arovas} D. P. Arovas and A. Auerbach, Phys. Rev. B {\bf 52}, 10114 (1995).
\bibitem{nersesyan} D. G. Shelton, A. A. Nersesyan, and A. M. Tsvelik,
Phys. Rev. B {\bf 53}, 8521 (1996)
\bibitem{sorensen} E.S. Sorensen, and I. Affleck, 
Phys. Rev. B {\bf 51}, 16115 (1995) 
\bibitem{zhang} S. Zhang, and D.P. Arovas, Phys. Rev. B {\bf 40}, 2708 (1989)
\end{thebibliography}
\end{document}